\documentclass[final]{appolb}
\usepackage{epsfig}

\begin{document}\hbadness=10000
\markboth{Jan Rafelski and Jean Letessier}
{Diagnosis of QGP with Strange Hadrons}
\title{\uppercase{Diagnosis of QGP with Strange Hadrons}}
\author{Jan RAFELSKI\footnote{Support by U.S. Department of Energy
under grant  DE-FG03-95ER40937\,.}
\address{Department of Physics, University of Arizona, 
Tucson, AZ 85721}
\and 
Jean LETESSIER
\address{Laboratoire de Physique Th\'eorique et Hautes 
Energies\footnote{LPTHE, Univ.\,Paris 6 et 7 is:
Unit\'e mixte de Recherche du CNRS, UMR7589.}\\
Universit\'e Paris 7, 2 place Jussieu, F--75251 Cedex 05}
}
\maketitle
\begin{abstract}
We review the current status of strangeness as signature of the
formation and dissociation of the deconfined QGP at the 
SPS energy scale, and present the status of our considerations
for  RHIC energies. By analyzing, within the framework of a
Fermi statistical model, the hadron abundance and spectra,
the properties of a disintegrating,  hadron evaporating,
deconfined QGP fireball are determined and can be compared with 
theory for the energy range 160--200$A$\,GeV 
on fixed target. We discuss in more detail our finding
that the pion yields occur near to pion condensation condition. 
Dynamical models of chemical strangeness equilibration are 
developed and applied to obtain strangeness production in
a QGP phase at conditions found at  SPS and expected at RHIC. 
 The sudden  QGP
break up model that works for the SPS data implies at RHIC
dominance of both baryon, and antibaryon, abundances 
by the strange baryon and antibaryon  yields.
\end{abstract}
\PACS{12.38.Mh, 25.75.-q, 25.75.Dw, 25.75.Ld}
\begin{center}
\vskip -15cm
{\bf Presented at the 29th Krak\'ow School of Theoretical Physics, 
Zakopane, June 1999 (Acta Physica Polonica B -- 1999/2000).}
\vskip 13.5cm
\end{center}

\section{Introduction}\label{introsec}
Strange particle signatures for the formation and evolution
of the deconfined quark-gluon phase of elementary matter 
(QGP) has been a subject developed quite intensely for the 
past 20 years. We review here our progress since 
the last major review \cite{acta96}, highlighting our 
analysis  of the Pb--Pb data at SPS, and  our 
predictions for hyperon yields from QGP at RHIC \cite{RL99}. 

We first describe in section~\ref{fermisec}, the 
Fermi model \cite{Fer50} analysis of the 
multi-particle  production processes in
158$A$ GeV Pb--Pb collisions carried out at CERN-SPS. 
Strongly interacting particles are believed to be 
produced with a probability commensurating to the 
size of the accessible phase space. 
The numerical methods which have been  developed in the 
context of an  analysis  of the 
lighter  200$A$ GeV S--Au/W/Pb system  \cite{LRa99}
are described.  We have in particular shown~\cite{LRPb98,LRprl99}
 that consideration of the light quark 
chemical non-equilibrium is necessary in order to arrive at
a  consistent interpretation of the experimental 
results of both the wide acceptance NA49-experiment
\cite{Ody98,Puh98,Bor97,App98,Mar99}
and central rapidity  strange (multi)strange (anti)baryon  
WA97-experiment \cite{Hol97,Kra98,WA97}.
This resulted also in considerable reduction of the 
chemical freeze-out temperature: we find $T_f=145\pm5$\,MeV,
while  originally it has been estimated to be 
\cite{LRT98,Bec98} $T_f=180$--$290$\,MeV.

Such a low freeze-out temperature is more consistent with 
the assumption we make that
there is no change of hadronic particle abundance after
the deconfined QGP source has dissociated. 
 This scheme is called {\it sudden hadronization} \cite{KMR86,Raf91}.
This can occur if  hadronic particles are 
produced either in:\\ 
\indent a) an evaporation process from a hot 
expanding,  surface or\\ 
\indent b) a sudden global hadronization process.

Our sudden hadronization scheme works very well, and
can be considered as established on view of many 
studies that could describe quite diverse data. 
Certain surprising features of Pb--Pb results 
that are seen within such analysis, and in particular
 the finding that the pion yield is governed by a fugacity that
is close to  the condensation point, as we 
shall show in  section~\ref{spsec}, lead us presently 
to favor the scenario b). That being the case
one may further suppose that a super-cooled plasma 
occupies a relatively large spatial volume, and 
it undergoes a global explosive decomposition
into individual hadrons, maximizing hadron 
occupancies and thus the entropy
content in the confined phase at the
near-pion-condensation condition. 

Pertinent  results of our analysis of the Pb--Pb system are
addressed in section~\ref{spsec}, where we have reevaluated 
our current results in consideration of some small change of 
the experimental data. We address all 
available SPS NA49 and WA97 experimental data, 
except for  $\Omega$ and $\overline\Omega$
particles. It is important to realize that if we succeed to describe well
a particle yield within the Fermi model, 
it means that the majority of all particles of the
particular type is produced by the statistical mechanisms we 
address here. In principle there could be many other production
mechanisms, and they add to the yields. Thus if our
description fail, an acceptable failure is the one which
{\it under-predicts the yield}. When statistical
model predicts very little if any production, the other reaction 
pictures may indeed be dominant and we should at least hesitate 
in our attempt to describe all the rare particle yields. 
The prime candidate for such consideration and omission from
statistical analysis is the totally strange $\Omega(sss)$ and its
antiparticle: they are 
triply strangeness suppressed and are very heavy
with $M_\Omega=1672$\,MeV, thus again significantly suppressed,
especially at low chemical freeze-out temperature. 
Their total statistical 
multiplicity is by a good distance smallest of all `stable' 
hadrons. Consequently, their production pattern is easily 
altered by, \eg, in source strangeness clustering. We have 
found  that if we use the results we obtain about the 
properties of the source \cite{LRprl99},
in order to compute 
the yields of $\Omega$ and $\overline\Omega$,
we invariably see that we obtain only  a fraction, 40--50\%, of all 
particles observed. Our preliminary conclusion is that we
should NOT explore these particles in the statistical 
production model.

Among results that we obtain  in section~\ref{spsec}
is  the, on a first sight, surprising 
overpopulation of the strangeness phase space occupancy. 
We explain how this can occur in   section~\ref{sprodsec},  
where the kinetic
theory for computation of the chemical strangeness
flavor abundance equilibration is presented.
We extend our past study of strangeness production at 
SPS conditions and show that, at the time of  QGP breakup at
RHIC energies, there is also in general full chemical equilibrium,
indeed that one can expect over-saturation of strangeness
flavor, just as at SPS. Our numerical 
study is based on  the dynamics of the phase space
occupancy rather than particle density,
 and we eliminate much of the dependence on the
dynamical flow effects by incorporating in the dynamics considered
the hypothesis of entropy  conserving  matter flow and evolution. 
We will make two assumptions of relevance for the
results we obtain:\\
$\bullet$ the kinetic (momentum distribution) 
equilibrium is reached faster than the chemical (abundance) 
equilibrium \cite{Shu92,Alam94};\\
$\bullet$ gluons  equilibrate chemically significantly 
faster than strangeness \cite{Won97}.\\
The first assumption  allows to 
study only the chemical abundances, rather than
the full momentum distribution, which simplifies greatly
the structure of the master equations; the second  assumption 
allows to focus after an initial time $\tau_0$ has passed 
on the evolution of strangeness population: $\tau_0$ 
is the time required for the development to near chemical
equilibrium of the  gluon population. 
As we shall see, the strange quark mass $m_{s}$ is 
the only undetermined parameter
that enters strangeness yield calculations. 
The  overpopulation of
the strangeness phase space, seen in SPS data 
arises for relatively small $m_{s}$(1GeV)$\simeq 200$\,MeV.

In the following section~\ref{rhicsec}, we use the experience 
we have with the SPS systems  and with the theoretical studies 
of strangeness production in QGP, in order to estimate the
strange particle production that is likely to occur at RHIC.
Some remarkable particle abundance results  arise, since during the 
 break-up of the QGP phase there is considerable
advantage for strangeness flavor to stick to baryons.
This can be easily understood considering that production
 of strange baryons over kaons  is favored
by the energy balance, \ie: $E(\Lambda+\pi)<E(\mbox{N+K})$. 
Since at RHIC  most hadrons 
produced are mesons, and baryons form just a small fraction of
all particles, initially we expect and will show, in section~\ref{rhicsec}, 
that hyperon production dominates  baryon production, \ie,
most baryons and antibaryons produced  will be strange.
A remarkable consequence of the sudden hadronization 
scenario is that this situation is maintained and thus hyperon dominance
should be observed at  RHIC. If indeed this prediction 
 is born out in the experiment, it will prove that
the there was formation of deconfined phase, followed 
by sudden hadronization.

We note that 
at  SPS energies described in section~\ref{spsec},
there is still an appreciable 
relative baryon abundance among all hadrons (about 15\%) 
and thus while hyperon dominance  begins to set in, 
there are (literally speaking)
still  some non-strange baryons left. With increasing
energy the yield of strange quark pairs
per baryon rises, and at the same time
the relative abundance of baryons among all hadrons diminishes, 
the relative population of non-strange baryons decreases 
rather rapidly and at RHIC energies the hyperons and/or antihyperons
 are the dominant strange particle fractions.  

We are not aware that other studies reported in literature 
about RHIC conditions have this remarkable result, see, \eg,
\cite{Dum99}. It is thus interesting to record the 
two major quantitative differences of the behavior
of deconfined matter we are considering:\\ 
$\bullet$ in QGP the particle density is high enough to
assure that the required  abundance of strangeness 
can be actually produced \cite{acta96,RM82,MS86,BCD95},
while in hadron phase it was shown that, even at SPS energy, this 
is not the case~\cite{KR85}.\\ 
$\bullet$ overpopulation of 
hadron phase space occupancies  occurs naturally when the
entropy rich QGP phase disintegrates into hadrons, 
which cannot be  expected in hadron based kinetic 
reactions. 

\section{Contemporary Fermi model of Hadron Production}
\label{fermisec}
We use 6 parameters to characterize the spectra 
and abundances of particles. Will describe these 
discussing their values, assuming a QGP source:\\
\indent 1) The strange quark fugacity $\lambda_{s}=1$ can be
obtained from the  requirement that strangeness balances \cite{Let93}:
\begin{eqnarray}
\langle N_{s}-N_{\bar s}\rangle=0\,.
\end{eqnarray}
 However, the Coulomb distortion
of the strange quark phase space plays an important role in the
understanding of this constraint for Pb--Pb collisions \cite{LRPb98}, 
leading to the Coulomb-deformed value $\lambda_{s}=1.10$\,, see also
Eq.\,(\ref{lamQ}).\\
\indent 2) Strange quark phase space occupancy 
$\gamma_{s}$ can be computed as we show in section~\ref{sprodsec}
 within the established kinetic theory
framework for strangeness production \cite{acta96,RM82}. For a rapidly
expanding system the production processes will lead to an
oversaturated  phase space with $\gamma_{s}>1$\,.\\
\indent 3) The equilibrium phase space occupancy of light quarks 
$\gamma_{q}$ is expected to exceed unity  significantly 
to accommodate the excess  entropy content in the plasma phase \cite{Let93}.
There is an interesting constraint that arises if hadronization is sudden in the
sense that  particles are produced at the same time, forming for pions a 
Bose gas. As we shall discuss at the end of this section, 
see Eq.\,(\ref{piBos}) this leads to an
upper limit:
\begin{eqnarray}\label{gammaqc}
\gamma_{q}<\gamma_{q}^c\equiv e^{m_\pi/2T}\,.
\end{eqnarray}
\indent 4) The  collective surface expansion velocity should remain below 
the relativistic sound velocity   \cite{acta96}: 
\begin{eqnarray}
v_c\le 1/\sqrt{3}.
\end{eqnarray}
\indent 5,6) If we assume that the stopping of the baryon number and energy 
is similar~\cite{acta96}, we know the energy per baryon content in the reactions 
and then the  equations of state produce a further constraint between
chemical freeze-out temperature $T_{f}$ and  light quark fugacity $\lambda_{q}$\, 
or equivalently, the baryochemical potential:
\begin{eqnarray}\label{muB}
\mu_B=3\,T_{f}\ln \lambda_{q}.
\end{eqnarray}

The difference between $\lambda_i$ and $\gamma_i$ is that, \eg, 
for strange and anti-strange quarks 
the same factor $\gamma_{s}$ 
applies, while the antiparticle fugacity is inverse of the particle
fugacity. The proper statistical physics foundation of  $\gamma_i$ 
is obtained considering the maximum entropy principle: it has
been determined that while the limit $\gamma_i\to1$ maximizes the specific
chemical entropy, this maximum is extremely shallow, indicating that 
a system with dynamically evolving volume will in general find more effective
 paths to increase entropy, than offered by the establishment of 
the absolute chemical equilibrium \cite{entro}. 

The abundances of the final state particles is most 
conveniently described
by considering the  phase space 
distribution of particles. 
The relative number of primary particles
freezing out from a source is obtained
noting that the  fugacity and phase space occupancy
 of a composite hadronic  particle is  expressed
by its constituents and that the probability to find all
$j$-components contained within  the $i$-th  emitted particle is:
\begin{equation}\label{abund}
N_i\propto e^{-E_i/T}\prod_{j\in i}\gamma_j\lambda_j,
\ \ \lambda_i=\prod_{j\in i}\lambda_j,
\ \  \gamma_i=\prod_{j\in i}\gamma_j.
\end{equation}

Taking the Laplace transform, we find, \eg,
for the strange sector, to the following partition function, like
expression:
\begin{eqnarray}
\ln{\cal Z}_{s} = { {V T^3} \over {2\pi^2} }
\hspace{-0.3cm}&&\left\{(\lambda_{s} \lambda_{q}^{-1} +
\lambda_{s}^{-1} \lambda_{q}) 
\gamma_{s} \gamma_{q} C^{\rm s}_{\rm M}
F_K +(\lambda_{s} \lambda_{q}^{2} +
\lambda_{s}^{-1} \lambda_{q}^{-2}) 
\gamma_{s}\gamma_{q}^{2} C^{\rm
s}_{\rm B} F_Y \right.\nonumber \\ 
&&\hspace{0.3cm}\left.+ (\lambda_{s}^2 \lambda_{q} +
\lambda_{s}^{-2} \lambda_{q}^{-1}) 
\gamma_{s}^2\gamma_{q} C^{\rm
s}_{\rm B}  F_\Xi + (\lambda_{s}^{3} + \lambda_{s}^{-3})
\gamma_{s}^3 C^{\rm s}_{\rm B} F_\Omega\right\}\ , 
\label{4a}
\end{eqnarray}
where the kaon, hyperon, cascade and omega degrees of freedom are
included. The phase space factors
$F_i$ of the strange particles are (with $g_i$ describing the statistical
degeneracy):
\begin{eqnarray}
F_i&=&\sum_j g_{i_j} W(m_{i_j}/T)\ .
\label{FSTR}
\end{eqnarray}
In the resonance sums $\sum_j$ all known strange hadrons should be
counted.  The function $W(x)=x^2K_2(x)$, where $K_2$ 
is the modified Bessel function,
arises from the phase-space integral of the
different particle distributions $f(\vec p)$.
It is important to remember that this expression does not describe the
properties of a gas of hadrons, thus it is not a partition function,
even if we give the Laplace transform of the phase space such formal semblance.

When the source of the particles is subject to flow, the Laplace transform 
that leads to the above expression is considerably more involved. 
the spectra and thus also multiplicities
of particles emitted are obtained replacing
the Boltzmann factor in Eq.\,(\ref{abund}) by \cite{Hei92}:
\begin{eqnarray}\label{abundflow}
e^{-E_i/T}&\to& \frac1{2\pi}\int d\Omega_v
  \gamma_c(1-\vec v_{\rm c}\cdot \vec p_i/E_i)
  e^{-{{\gamma_cE_i}\over T}
    \left(1-\vec v_{\rm c}\cdot \vec p_i/E_i\right)},\nonumber\\
\gamma_c&=&\frac1{\sqrt{1-\vec v_{\rm c}^{\,2}}}\,,
\end{eqnarray}
a result which can  be intuitively obtained by a Lorentz
transformation between an observer on the surface of
the fireball, and one at rest in laboratory frame. In
certain details the results we obtain
confirm the applicability  of this simple approach.
We consider for SPS energy range
 the radial flow model, perhaps the  simplest of
the complex flow cases possible ,
but it suffices to fully assess the impact of flow on our analysis.

While the integral over  the entire phase space of the flow spectrum
yields as many particles with and without flow, when acceptance cuts are
present particles of different mass experience differing flow effects.
Here, we note that the final particle abundances measured in an experiment 
are obtained after all unstable hadronic resonances
 are allowed to disintegrate and feed the stable hadron spectra.
In order to  minimize the impact of unknown  flow pattern at hadron freeze-out 
when considering particle abundances measured
in a restricted phase space domain, we study particle abundance ratios 
involving what we call compatible hadrons: 
these are particles likely to be impacted in a similar fashion by 
collective flow dynamics in the fireball.

\begin{figure}[tb]
\vspace*{-1.6cm}
\centerline{
\psfig{width=8.5cm,figure=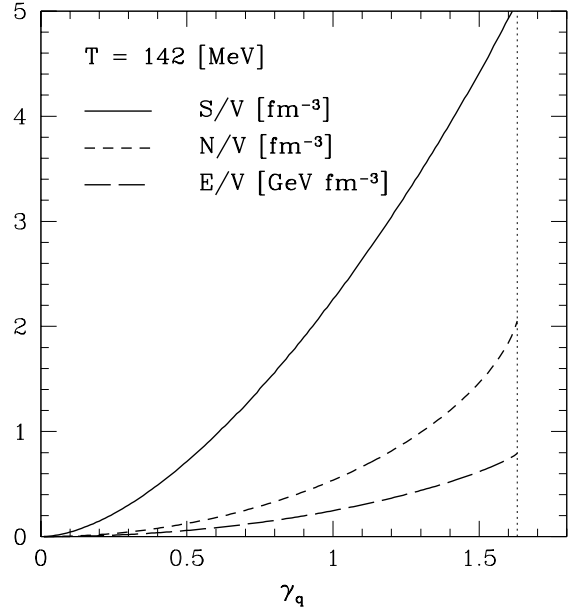}}
\vspace*{-2cm}
\caption{ 
Dependence of pion gas properties $N/V$-particle, $E/V$-energy and 
$S/V$-entropy  density, as function of  $\gamma_{q}$ at $T=142$\,MeV. 
\label{abssne}} 
\end{figure}

We now return to review the case of pions, which is exceptional since
we will be considering a rather large values of $\gamma_{q}>1.5$. 
The chemical fugacity for a particle composed of a light quark-antiquark pair
is $\gamma_{q}^2$. Thus the Bose  distribution in momentum space has the shape:
\begin{eqnarray}\label{piBos}
f_\pi(E)=\frac{1}{\gamma_{q}^{-2}e^{E_\pi/T}-1}\,,\qquad E_\pi=\sqrt{m_\pi^2+p^2}\,.
\end{eqnarray}
The range of values for $\gamma_{q}$ is bounded from above by the Bose
singularity. When $\gamma_{q}\to \gamma_{q}^c$, see Eq.\,(\ref{gammaqc}),
the lowest energy state (in the continuum limit with $p\to 0$ )
will acquire macroscopic  occupation and a pion condensate is formed. 
Such a condensate `consumes' energy without
consuming entropy of the primordial high entropy QGP phase. 
Thus a condensate is not likely to develop, but the hadronization
process may have the tendency to approach the limiting value in
order to more efficiently connect the deconfined and 
the confined phases, since, as we show in Fig.\,\ref{abssne},
the entropy density is nearly twice as high at $\gamma_{q}\simeq 
\gamma_{q}^c$ than at $\gamma_q=1$. 

\begin{figure}[t]
\vspace*{-1.3cm}
\centerline{ 
\psfig{width=6.8cm,figure=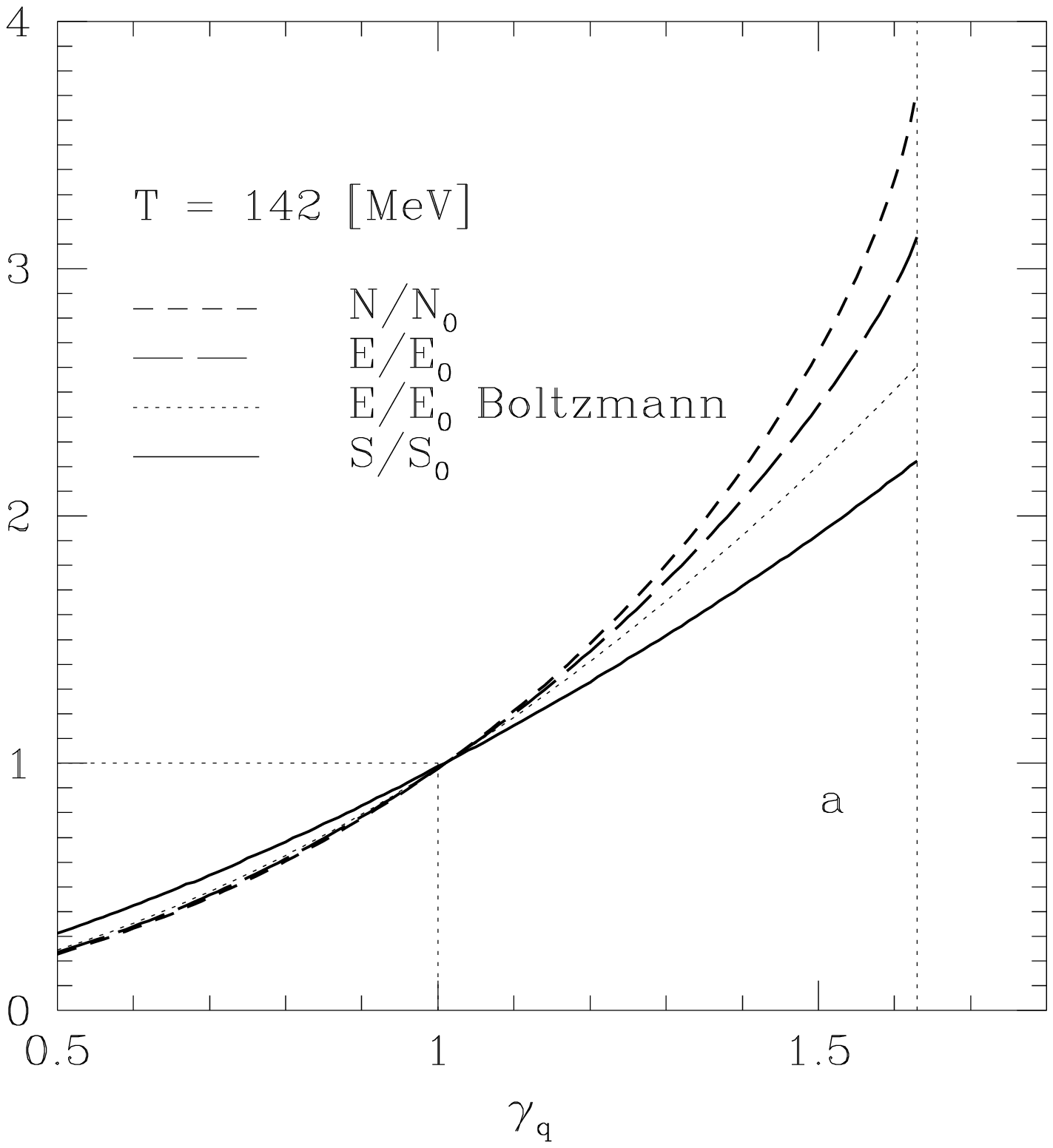}\hspace*{-0.5cm}
\psfig{width=6.8cm,figure=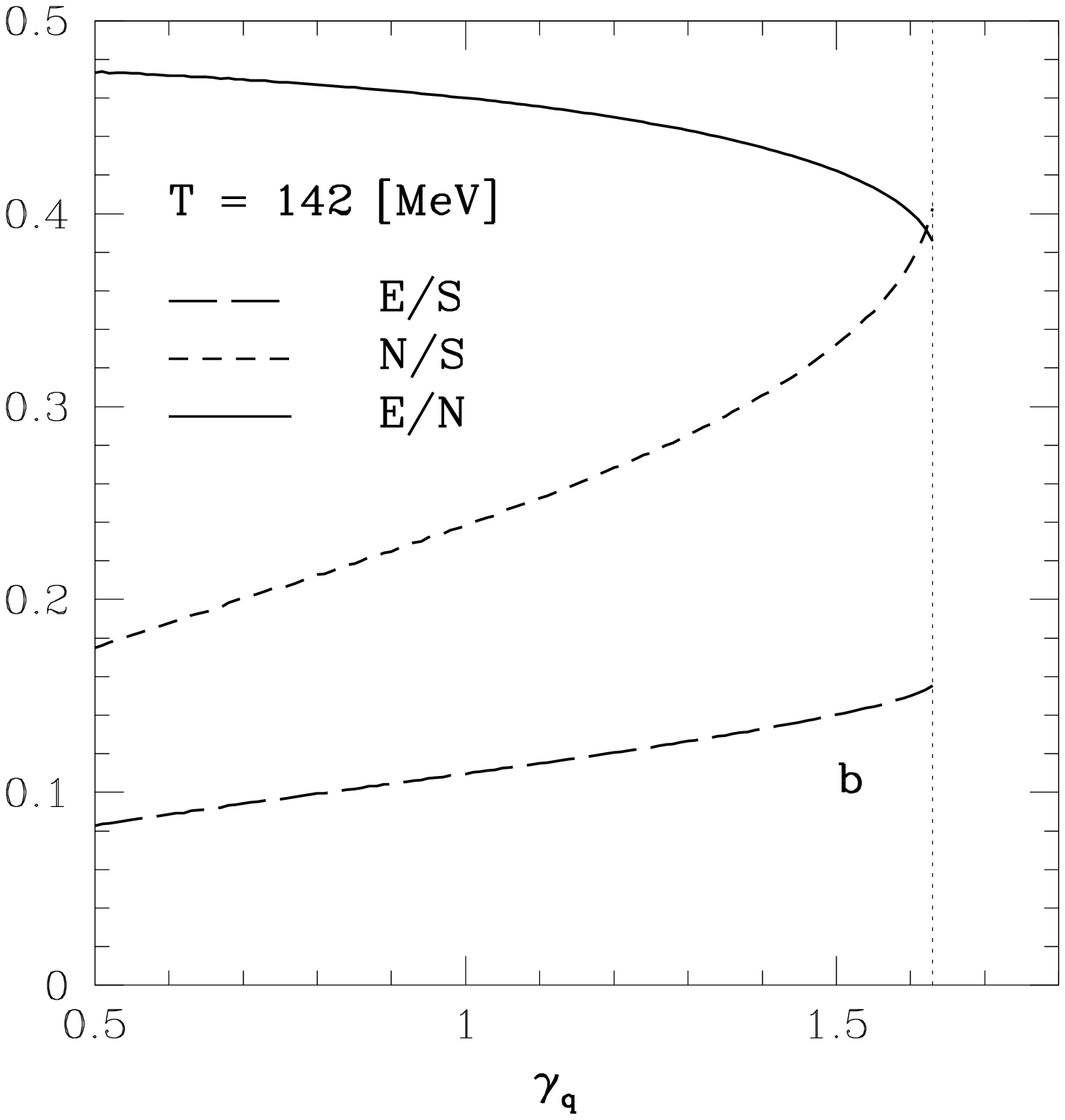}}
\vspace*{-1.5cm}
\caption{ 
Dependence of pion gas properties ($N$-particle, $E$-energy and 
$S$-entropy)  density 
as function of  $\gamma_{q}$ for $T=142$\,MeV. 
{\bf a)} ratios relative to equilibrium value $\gamma_q=1$; 
{\bf b)} relative ratios, thus $E/N$, $S/N$ and $E/S$.
\label{ratiosne}} 
\end{figure}

 To see clearly how this can occur, 
 we looked more closely at the relative properties of 
a pion gas for $\gamma_{q}\to \gamma_{q}^c$\,.
In Fig.\,\ref{ratiosne}, we see the relative change
in energy per pion,  (inverse of) entropy per pion, and
energy per entropy, for 
fixed $T=142$\,MeV corresponding to our best fit condition.
We see that a hadronizing gas will consume at higher $\gamma_q$ less
energy per particle, and that the energy per entropy is nearly
constant. Dissociation into pions at  $\gamma_{q}\to \gamma_{q}^c$
appears thus to be an effective way to convert excess of entropy 
in the plasma into hadrons, without need for reheating, or a mixed
phase which would allow the volume to grow. In short, the finding of 
the maximum allowable $\gamma_{q}$ is intrinsically consistent with
the notion of an explosively disintegrating QGP phase.

\section{Update of SPS Experimental Data Analysis}
\label{spsec}
In the past year our  work addressed our discovery that
consideration of the light quark 
chemical non-equilibrium is necessary in order to arrive at
a  consistent interpretation of the experimental 
results emanating from CERN \cite{PRL99}. We have also 
incorporated in our earlier analysis 
of the Pb--Pb system \cite{LRPb98} a study 
of collective matter flow. Properties of the dense 
fireball as determined in this approach offer clear
 evidence that a QGP disintegrates  at $T_f\simeq$\,144\,MeV,
corresponding to energy density 
$\varepsilon=\cal O$(0.5) GeV/fm$^3$ \cite{Kar98}. 
With flow, the  analysis addresses also the 
 $m_\bot$-slopes of strange particles. Notably, 
the near equality of (inverse) slopes of nearly all strange
baryons and antibaryons  arises by means of the  
sudden hadronization at the surface of an
exploding QGP fireball. In the hadron based microscopic 
simulations this behavior of  $m_\bot$-slopes can also
arise allowing for particle-dependent freeze-out times \cite{HSX98}.

We note that though we use six parameters to characterize the hadron 
phase space at chemical freeze-out, compare section \ref{fermisec}, 
there are only two truly 
unknown properties: the chemical freeze-out temperature $T_{f}$ and  
the light quark fugacity $\lambda_{q}$\, 
(or equivalently, the baryochemical potential Eq.\,(\ref{muB}))
--- we recall that the parameters 
$\gamma_i,\,i=q,s$ controls overall
abundance of  quark pairs, while $\lambda_i$ controls the difference
between quarks and anti-quarks of given flavor.  As already noted 
earlier, the  four other
parameters are not arbitrary, and we could have used their
tacit and/or computed values:\\
\indent 1) the strange quark fugacity $\lambda_{s}$ is usually
fixed by the  requirement that strangeness balances 
$\langle s-{\bar s}\rangle=0$\, \cite{Raf91}. The Coulomb distortion
of the  strange quark phase space plays an important role in the
understanding of this constraint for Pb--Pb collisions,
see Eq.\,(\ref{lamQ}) \cite{LRPb98};\\
\indent  2) the strange quark phase space occupancy 
$\gamma_{s}$ can be computed  within the established kinetic theory
framework for strangeness production \cite{acta96,RM82};\\
\indent  3) the tacitly assumed equilibrium phase space occupancy of light quarks 
$\gamma_{q}=1$\,;\\
\indent  4) assumed collective expansion to proceed at
the relativistic sound velocity, $v_c=1/\sqrt{3}$ \cite{acta96}.\\
However, the rich particle data basis allows us to find from experiment
the actual values of these four parameters, allowing to confront
the theoretical results and/or hypothesis with experiment.

The value of $\lambda_{s}$ we obtain from the strangeness conservation 
condition $\langle s-{\bar s}\rangle=0$\ in QGP is, to a very good 
approximation \cite{LRPb98}:
\begin{equation}\label{tilams}\label{lamQ}
\tilde\lambda_{s}\equiv \lambda_{s} \lambda_{\rm Q}^{1/3}=1\,,\qquad
\lambda_{\rm Q}\equiv
\frac{\int_{R_{\rm f}} d^3r e^{\frac V{T}} } {\int_{R_{\rm f}} d^3r}\,.
\end{equation}
 $\lambda_{\rm Q}<1$  expresses the Coulomb deformation of 
strange quark phase space. This effect is
relevant in central Pb--Pb interactions, but not in 
S--Au/W/Pb reactions. $\lambda_{\rm Q}$ is not a fugacity that 
can be adjusted to satisfy a chemical condition,
since consideration of $\lambda_i,\ i=u,d,s$ exhausts all available
chemical balance conditions for the abundances of hadronic particles. 
The subscript ${R_{f}}$ in Eq.\,(\ref{lamQ}) reminds us 
that the classically
allowed region within the dense matter fireball is included in
the integration over the  level density.
 Choosing $R_{\rm f}=8$\,fm, $T=140$\,MeV,
$m_{s}=200$\,MeV (value of $\gamma_{s}$ is practically irrelevant),
for $Z_{\rm f}=150$ the value is $\lambda_{s}=1.10$\,.

The available compatible particle yield ratios (excluding 
$\Omega$ and $\overline\Omega$, see section~\ref{introsec})
 are listed  in table~\ref{resultpb2}, top section from the
experiment WA97 for $p_\bot>0.7$ GeV within a narrow
$\Delta y=0.5$ central rapidity window. Further below 
are shown  results from the large  acceptance experiment NA49, 
extrapolated to full $4\pi$ phase space coverage. 
We first fit 11 experimental results shown in table~\ref{resultpb2},
and than turn to include also the $m_\bot$-slope  in our considerations,
and thus have 12 data points.  The total error: 
\begin{equation}\label{chi2}
\chi^2_{\rm T}\equiv\frac{\sum_j({R_{\rm th}^j-R_{\rm exp}^j})^2}{
({{\Delta R _{\rm exp}^j}})^2}
\end{equation}
for the  four theoretical 
columns is shown at the bottom of this table
along with the number of data points `$N$', parameters `$p$' 
used and  (algebraic) redundancies `$r$' connecting the 
experimental results. For $r\ne 0$ it is more appropriate 
to quote the total  $\chi^2_{\rm T}$, with a initial qualitative
statistical relevance condition  $\chi^2_{\rm T}/(N-p)<1$.
The first theoretical columns refer to results without
collective velocity $v_c$ (subscript $0$) the three other 
with fitted $v_c$ (subscript $v_c$). In column three, superscript `$sb$' 
means that $\lambda_{s}$ is fixed by strangeness balance and, in column 
four, superscript `$sc$' means that $\gamma_{q}=\gamma_{q}^c=e^{m_\pi/2T_f}$, 
that is $\gamma_{q}$ is fixed by its upper limit, the pion condensation point. 
All results have been newly 
recomputed, to account for slightly higher value of the
ratio  $h^-/B$ \cite{App99}.

\begin{table}[t]
\caption{\label{resultpb2}
WA97 (top) and NA49 (bottom)  Pb--Pb 158$A$ GeV particle ratios
and some of our theoretical results, see text for explanation.}
\begin{center}
\begin{tabular}{lcl|l|lll}
\hline\hline
 Ratios & $\!\!\!\!$Ref. &  Exp.Data                           &Pb$|_0$& Pb$|_v$ & Pb$|_v^{\rm sb}$ & Pb$|_v^{\rm sc}$  \\
\hline
${\Xi}/{\Lambda}$ &\cite{Kra98} &0.099 $\pm$ 0.008                    &  0.104  & 0.103  & 0.105 & 0.103\\
${\overline{\Xi}}/{\bar\Lambda}$ &\cite{Kra98} &0.203 $\pm$ 0.024     &  0.214  & 0.208  & 0.209 & 0.206\\
${\bar\Lambda}/{\Lambda}$  &\cite{Kra98} &0.124 $\pm$ 0.013           &  0.124  & 0.125  & 0.124 & 0.125\\
${\overline{\Xi}}/{\Xi}$  &\cite{Kra98} &0.255 $\pm$ 0.025            &  0.256  & 0.252  & 0.248 & 0.251\\
\hline
$(\Xi+\bar{\Xi})\over(\Lambda+\bar{\Lambda})$&\cite{Ody97}  &0.13 $\pm$ 0.03
                                                                      &  0.126  & 0.122  & 0.124 & 0.122\\
${K^0_{s}}/\phi$   &\cite{Puh98}  & 11.9 $\pm$ 1.5\ \             &  14.2   & 13.3   & 13.0  & 13.4 \\
${K^+}/{K^-}$         &\cite{Bor97}         &  1.80$\pm$ 0.10         &  1.80   & 1.82   & 1.78  & 1.83 \\
$p/{\bar p}$     &\cite{Ody98}              &18.1 $\pm$4.\ \ \ \      &  17.3   & 16.7   & 16.6  & 16.6 \\
${\bar\Lambda}/{\bar p}$     &\cite{Roh97}  & 3. $\pm$ 1.             &  2.68   & 2.11   & 2.11  & 2.11 \\
${K^0_{s}}$/B       &\cite{Jon96}       & 0.183 $\pm$ 0.027       &  0.181  & 0.181  & 0.163 & 0.188\\
${h^-}$/B                 &\cite{App99}     & 1.97 $\pm $ 0.1\ \      &  1.96   & 1.97   & 1.97  & 1.96 \\
\hline
 & $\chi^2_{\rm T}$     &                                             &  3.6    & 2.5    & 3.2   & 2.6  \\
 &  $ N;p;r$     &                                                    & 11;5;2  & 12;6;2 & 12;5;2& 12;5;2\\
\end{tabular}
\end{center}
\vskip -0.8cm
\end{table}

It is interesting to note that the highest confidence result is 
obtained in the last column, just when the light quark 
phase space occupancy assumes value at the pion condensation point: 
here the number of degrees of freedom is higher than in the
second column, obtained without constraint. It is unclear
at present what is the full extent of this remarkable 
result. Another interesting insight is that radial flow 
always on its own improves  our ability to describe the 
data. However, $m_\bot$ spectra offer another independent measure 
of flow, and confirm  very strongly our findings about the value
of $v_c$. We proceeded as follows: for a given pair of values
 $T_{f}$ and $v_{\rm c}$ we evaluate the resulting
$m_\bot$ particle spectrum and analyze it using the spectral shape
and kinematic cuts employed by the experimental groups.
To find the best values we consider just one `mean' strange baryon
experimental  value ${\bar T}_{\bot}^{\rm Pb}=260\pm10$,
since within the error the high $m_\bot$ strange (anti)baryon
inverse slopes are overlapping. Thus when considering $v_c$ along with
${\bar T}_{\bot}$ we have one parameter and one data point more. 
Once we find best values of $T_{\rm f}$
and $v_{\rm c}$, we  study again the inverse slopes of 
individual  particle spectra. We obtain an acceptable
agreement with the experimental $T_{\bot}^j$ as 
shown in left section of table~\ref{Tetrange2}\,.

For comparison, we have also considered in the same framework the
S-induced reactions, and the right section of 
table~\ref{Tetrange2} shows a  good
agreement with the WA85 experimental data \cite{WA85slopes}.
We used as the `mean' experimental slope data point
${\bar T}_{\bot}^{\rm S}=235\pm10$. We can see that
within a significantly smaller error bar, we obtained an accurate
description of the $m_\bot^{\rm S}$-slope data. This analysis 
implies that  the kinetic freeze-out, where elastic particle-particle 
collisions cease, cannot be occurring at a condition very different 
from the chemical freeze-out.  However, one pion
HBT analysis at $p_\bot<0.5$ GeV suggests
kinetic pion freeze-out at about $T_k\simeq120$ MeV \cite{Fer99}.
A possible explanation of why here considered $p_\bot>0.7$ GeV
particles are not subject to a greater spectral deformation
after chemical freeze-out, is  that they escape before the bulk 
of softer hadronic particles is formed. At least for strange 
brayons and antibaryons this is the result also seen in 
a recent microscopic study of the freeze-out process \cite{Dum99}.
\begin{table}[tb]
\caption{\label{Tetrange2}
Experimental and theoretical $m_\bot$ spectra inverse slopes $T_{\rm th}$.
Left Pb--Pb results 
from experiment NA49 \protect\cite{Mar99}
for  kaons and from experiment WA97 \protect\cite{WA97} for baryons;
right  S--W  results from  WA85  \protect\cite{WA85slopes}.}
\begin{center}
\begin{tabular}{l|cc|cc}
\hline\hline
               & $T_{\bot}^{\rm Pb}$\,[MeV]&$T_{\rm th}^{\rm Pb}$\,[MeV]&$T_{\bot}^{\rm S}$\ [MeV]&
                                                                   $T_{\rm th}^{\rm S}$\ [MeV]\\
\hline
$T^{{\rm K}^0}$             & 223 $\pm$  13&  241& 219 $\pm$  \phantom{1}5 &  215\\
$T^\Lambda$                 & 291 $\pm$  18&  280& 233 $\pm$  \phantom{1}3 & 236\\
$T^{\overline\Lambda}$      & 280 $\pm$  20&  280& 232 $\pm$  \phantom{1}7 & 236\\
$T^\Xi$                     & 289 $\pm$  12&  298& 244 $\pm$  12& 246\\
$T^{\overline\Xi}$          & 269 $\pm$  22&  298& 238 $\pm$  16& 246\\
\end{tabular}
\end{center}
\vskip -0.8cm
\end{table}

The six statistical parameters describing the  particle abundances
are shown in the top section of table~\ref{fitqpbs}, where we
also show in the last column for comparison, the best result for S-induced
reactions, where the target has been W/Au/Pb \cite{LRa99}.  
The  errors in the results are one standard  deviation errors arising
from the propagation of the experimental measurement
error, but apply only when the  theoretical model describes 
the data well. All results shown  in 
table~\ref{fitqpbs} have convincing statistical confidence level.
For the S-induced reactions the number 
of redundancies $r$ shown in heading of the table~\ref{fitqpbs}
is large, since same data comprising
different kinematic cuts has been included in the analysis. 
It is quite reassuring that within error the freeze-out 
temperature $T_{\rm f}$ seen in table~\ref{fitqpbs},
is the same for both the S- and Pb-induced reactions,
even though the chemical phase space occupancies differ greatly. 
This must be the case within our model of sudden freeze-out
and constitutes its firm confirmation. The variation in the 
shape of the particle spectra  is fully explained by a change in
the collective velocity, which  rises from
$v_c^{\rm S}=0.49\pm0.02$ to $v_c^{\rm Pb}=0.54\pm0.04\simeq 1/\sqrt{3}=0.577$.
The light quark fugacity $\lambda_{q}$ implies that baryochemical
potential is 
$\mu_B^{\rm Pb}=203\pm5 >\mu_B^{\rm S}=178\pm5$\,MeV.
As in  S-induced reactions where $\lambda_{s}=1$,
now  in Pb-induced reactions, a value $\lambda_{s}^{\rm Pb}\simeq 1.1$
characteristic for a source of freely movable  strange quarks with
balancing strangeness, \ie, $\tilde\lambda_{s}=1$ is obtained, 
see Eq.\,(\ref{lamQ}). 
\begin{table}[tb]
\caption{\label{fitqpbs}
Heading: $\chi^2_{\rm T}$, number of data points $N$, parameters $p$ and 
redundancies $r$;  upper section: statistical model parameters
which best describe the experimental results for
Pb--Pb data, and in last column for S--Au/W/Pb data presented in 
Ref.\,\protect\cite{LRa99}\,.
Bottom section: specific energy, entropy, anti-strangeness, net strangeness
 of  the full hadron phase space characterized by these
statistical parameters. In column three we fix $\lambda_{s}$ by requirement of 
strangeness conservation, and in column four we choose $\gamma_{q}=\gamma_{q}^c$, the pion condensation point.}
\vspace{-0.2cm}\begin{center}
\begin{tabular}{l|ccc|c}
\hline\hline
                       & Pb$|_v$            & Pb$|_v^{\rm sb}$ & Pb$|_v^{\rm sc}$        & S$|_v$ \\
$\chi^2_{\rm T};\ N;p;r$&2.5;\ 12;\,6;\,2   & 3.2;\ 12;\,5;\,2 & 2.6;\ 12;\,5;\,2        &  6.2;\ 16;\,6;\,6 \\
\hline
$T_{f}$ [MeV]          &    142 $\pm$ 3     &  144 $\pm$ 2     &  142 $\pm$ 2            &  144 $\pm$ 2 \\
$v_c$                  &   0.54 $\pm$ 0.04  & 0.54 $\pm$ 0.025 & 0.54 $\pm$ 0.025        &   0.49 $\pm$ 0.02\\
$\lambda_{q}$          &   1.61 $\pm$ 0.02  & 1.605 $\pm$ 0.025& 1.615 $\pm$ 0.025       &   1.51 $\pm$ 0.02 \\
$\lambda_{s}$          &   1.09 $\pm$ 0.02  & 1.10$^*$         & 1.09 $\pm$ 0.02         & 1.00 $\pm$ 0.02   \\
$\gamma_{q}$           &   1.7 $\pm$ 0.5    & 1.8$\pm$ 0.2   &$\gamma_{q}^c=e^{m_\pi/2T_f}$&   1.41 $\pm$ 0.08 \\
$\gamma_{s}/\gamma_{q}$&   0.79 $\pm$ 0.05  & 0.80 $\pm$ 0.05  & 0.79 $\pm$ 0.05         &  0.69 $\pm$ 0.03  \\
\hline
$E_{f}/B$              &   7.8 $\pm$ 0.5    & 7.7 $\pm$ 0.5    & 7.8 $\pm$ 0.5           &  8.2 $\pm$ 0.5    \\
$S_{f}/B$              &    42 $\pm$ 3      & 41 $\pm$ 3       & 43 $\pm$ 3              &   44 $\pm$ 3     \\
${s}_{f}/B$            &  0.69 $\pm$ 0.04   & 0.67 $\pm$ 0.05  & 0.70 $\pm$ 0.05         &   0.73 $\pm$ 0.05 \\
$({\bar s}_f-s_f)/B$   &  0.03 $\pm$ 0.04   & 0$^*$            &  0.04 $\pm$ 0.05        &    0.17 $\pm$ 0.05\\
\end{tabular}
\end{center}
\vskip -0.8cm
\end{table}

The values of $\gamma_{q}>1$, seen in table~\ref{fitqpbs}, imply  that there is
phase space over-abundance of light quarks, to which,
\eg, gluon fragmentation at QGP breakup {\it prior} to hadron 
formation contributes.  $\gamma_{q}$ assumes in our data 
analysis a value near to  where pions 
could begin to  condense \cite{PRL99}, Eq.\,(\ref{gammaqc}). 
We found studying the ratio $h^-/B$
separately from other experimental results
that the value of $\gamma_{q}\simeq\gamma_{q}^c$ is fixed
consistently and independently both, by the negative hadron ($h^-$),
and the strange hadron yields. The unphysical  range 
$\gamma_{q}>\gamma_{q}^c\simeq 1.63$ can  arise 
(see column Pb$\vert_v^{\mbox sb}$)  since, up to this
point, we use only a first quantum (Bose/Fermi) 
correction. However, when  Bose distribution for 
pions is implemented, which requires  the 
constraint $\gamma_{q}\le\gamma_{q}^c$, 
we obtain  practically the same results, as shown
in second column of table~\ref{fitqpbs}. 
We then show in table~\ref{fitqpbs} the ratio $\gamma_{s}/\gamma_{q}\simeq 0.8$,
which corresponds (approximately)  to the parameter $\gamma_{s}$ when
$\gamma_{q}=1$ had been  assumed.  We note  that $\gamma_{s}^{\rm Pb}>1$.
This strangeness over-saturation effect could arise from the effect 
of gluon fragmentation combined with early chemical equilibration
in QGP, $\gamma_{s}(t<t_f)\simeq 1$. The ensuing rapid expansion
preserves this high strangeness yield, and thus we find the result
$\gamma_{s}>1$\,, as is shown in Fig.\,33 in \cite{acta96}.

We show, in the bottom section of table~\ref{fitqpbs}, the
energy and entropy content per baryon,
and specific anti-strangeness content,
along with specific strangeness asymmetry 
of the hadronic particles emitted.
The  energy  per baryon seen in the emitted hadrons is nearly
equal to the available specific energy
of the collision  (8.6 GeV for Pb--Pb, 8.8--9 GeV for S--Au/W/Pb).
This implies that the fraction of energy deposited in the central
fireball  must be nearly the same as the fraction of baryon number.
The small reduction of the specific entropy in Pb--Pb compared to
the lighter S--Au/W/Pb system maybe driven by the greater baryon
stopping in the larger system, also seen in the smaller energy per
baryon content. Both collision systems freeze out at energy per unit
of entropy $E/S=0.185$ GeV. 
There is a loose relation of this universality in the 
chemical freeze-out condition with the suggestion made
recently that particle freeze-out occurs at a fixed energy per baryon for
all physical systems \cite{CR98}, since the entropy content is related to
particle multiplicity. The overall high specific entropy content we find
agrees well with the entropy content
evaluation made  earlier \cite{Let93} for the S--W case.

Inspecting Fig.\,38 in \cite{acta96}, we see
that the specific yield of strangeness we expect
from the kinetic theory in QGP is at the level of 0.75 
per baryon, in agreement with the results of
present analysis shown in table \ref{fitqpbs}. 
This high strangeness yield leads
to the enhancement of multi-strange (anti)baryons,
which are viewed as important hadronic signals of
QGP phenomena \cite{Raf80}, and a series of recent
experimental analysis has carefully demonstrated  comparing
p--A with A--A results that
there is quite significant enhancement \cite{WA97,WA85}, 
as has also been noted before by the experiment NA35 \cite{Alb94}.
The strangeness imbalance seen in the asymmetrical S--Au/W/Pb system
(bottom of table  \ref{fitqpbs}) could be a real effect arising from 
hadron phase space properties. However, this result
also reminds us that though the statistical errors are very small, 
there could be in this asymmetric system a considerable
systematic error due to presence of a significant spectator 
matter component. In the symmetric Pb--Pb collisions this 
effect disappears, despite the fact that the freeze-out flow 
pattern could be much more complex and
there could be a distortion of particle spectra 
at low momenta not accounted for in our study, for we 
did not model the ratio of kaons to hyperons. Considering 
this limitation  it is  indeed remarkable, how well 
the conservation of strangeness
condition  is satisfied, when it is not being enforced. 

\section{RHIC and Dynamics of Strangeness Production}
\label{sprodsec}
In some key aspects, the methods to describe strangeness
production which  we have been developing
differ from those obtained in other studies of chemical 
equilibration of quark flavor, in particular for RHIC 
conditions \cite{Bir93,Won96,Sri97}. 
For example,   we use  running  QCD parameters (both 
coupling and strange quark mass) with strong 
coupling constant $\alpha_{s}$  as determined 
 at the $M_{Z^0}$ energy scale. We also incorporate entropy
conserving flow into the dynamical equations directly,
exploiting significant cancellations that occur, and
thus obtain a relatively simple dynamical model for the 
evolution of the phase space occupancy $\gamma_{s}$ 
of strange quarks in the expanding QGP. 

The phase space  distribution  $f_{s}$  
can be  characterized by a local 
temperature $T(\vec x,t)$ of a (Boltzmann) equilibrium distribution  
$f_{s}^\infty$\,, with  normalization set 
by a phase space occupancy factor:
\begin{equation}\label{gamdef2}
f_{s}(\vec p,\vec x; t))\simeq \gamma_{s}(T) 
   f_{s}^\infty(\vec p;T)\,.
\end{equation}
Eq.\,(\ref{gamdef2})  invokes in the momentum independence of 
$\gamma_{s}$ our  first assumption. More generally,
the factor $\gamma_i,\, i=g,q,s,c$, allows a local density 
of gluons, light quarks, strange quarks and charmed quarks,
respectively not to be 
determined by the local momentum shape, but to evolve 
independently.  With variables 
$(t,\vec x)$ referring to an observer in the laboratory 
frame, the chemical evolution can be described by the strange
quark  current non-conservation arising from strange quark 
pair production described by a Boltzmann collision term:
\begin{eqnarray}
\partial_\mu j^\mu_{s}\equiv  {\partial \rho_{s}\over \partial t} +
   \frac{ \partial \vec v \rho_{s}}{ \partial \vec x}
&\!=& \!\frac12
\rho_g^2(t)\,\langle\sigma v\rangle_T^{gg\to s\bar s}\nonumber\\  
&+&\!\rho_{q}(t)\rho_{\bar q}(t)
\langle\sigma v\rangle_T^{q\bar q\to s\bar s}\!\! -\!
\rho_{s}(t)\,\rho_{\bar{\rm s}}(t)\,
\langle\sigma v\rangle_T^{s\bar s\to gg,q\bar q}\!.
\label{drho/dt1}
\end{eqnarray}
The factor 1/2  avoids double counting of gluon pairs.
The implicit sums over spin, color and any other 
discreet quantum numbers are combined in the particle density 
 $\rho=\sum_{s,c,\ldots}\int d^3p\,f$, and  we have also 
introduced the  momentum averaged production/annihilation
thermal reactivities (also called `rate coefficients'):
\begin{equation}
\langle\sigma v_{\rm rel}\rangle_T\equiv
\frac{\int d^3p_1\int d^3p_2 \sigma_{12} v_{12}f(\vec p_1,T)f(\vec p_2,T)}
{\int d^3p_1\int d^3p_2 f(\vec p_1,T)f(\vec p_2,T)}\,.
\end{equation}
$f(\vec p_i,T)$ are the relativistic Boltzmann/J\"uttner
distributions of two colliding particles $i=1,2$ of momentum $p_i$. 

The  current conservation used above in the laboratory
`Eulerian' formulation  can also be written with reference to the 
individual particle dynamics in the so called `Lagrangian' 
description: consider
$\rho_{s}$  as the  inverse of the small
volume available to each particle. Such a volume is defined 
in the local frame of reference for which 
the local flow vector vanishes $\vec v(\vec x,t)|_{\mbox{local}}=0$.
The considered volume $\delta V_l$ being 
occupied by small number of particles 
$\delta N$ (\eg, $\delta N=1$), we have:
  \begin{equation}\label{Nsinf}
\delta N_{s}\equiv \rho_{s} \delta V_l \,.
\end{equation}
 The left hand side (LHS) 
of Eq.\,(\ref{drho/dt1}) can be now written as: 
\begin{equation}\label{LagCor}
{\partial \rho_{s}\over \partial t} +
   \frac{ \partial \vec v \rho_{s}}{ \partial \vec x}\equiv
       \frac{1}{\delta V_l}\frac{d\delta N_{s}}{dt}= 
\frac{d\rho_{s}}{dt}+\rho_{s}\frac1{\delta V_l}\frac{d\delta V_l}{dt}\,.
\end{equation}
Since $\delta N$ and $\delta V_l dt$ are L(orentz)-invariant, 
the actual choice of the frame of reference in which the 
right hand side (RHS) of Eq.\,(\ref{LagCor}) is 
studied is irrelevant and we drop henceforth the subscript $l$.

We can further adapt Eq.\,(\ref{LagCor}) to the dynamics we pursue:
we introduce  
$\rho_{s}^\infty(T)$ as the (local) chemical equilibrium  abundance 
of strange quarks, thus $\rho=\gamma_{s}\rho_{s}^\infty$.
We evaluate the equilibrium abundance  
$\delta N_{s}^\infty=\delta V\rho_{s}^\infty(T)$
 integrating the Boltzmann distribution:
\begin{equation}\label{Nsinfty}
\delta N_{s}^\infty=[\delta VT^3] {3\over\pi^2} \,z^2K_2(z)\,,
\quad z={m_{s}\over T}\,,
\end{equation}
where $K_\nu$ is the modified Bessel function of order $\nu$; we 
will below use: $d[z^\nu K_\nu(z)]/dz=-z^\nu K_{\nu-1}$\,.
The first factor on the RHS
 in Eq.\,(\ref{Nsinfty}) is a constant in time should the
evolution of matter after the
initial pre-thermal time period $\tau_0$ 
be entropy conserving \cite{Bjo83}, and thus 
 $\delta VT^3=\delta V_0T^3_0$=Const.\,.
We now substitute in Eq.\,(\ref{LagCor}) 
and obtain
\begin{equation}\label{lochem}
{\partial \rho_{s}\over \partial t} +
   \frac{ \partial \vec v \rho_{s}}{ \partial \vec x}=  
\dot T\rho_{s}^\infty\left({{d\gamma_{s}}\over{dT}}+
\frac{\gamma_{s}}{T}z\frac{K_1(z)}{K_2(z)}\right)\,,
\end{equation}
where $\dot T=dT/dt$. Note that in Eq.\,(\ref{lochem}) 
only  a part of  the usual flow-dilution term
is left, since we implemented the adiabatic volume expansion,
and study the evolution of
the phase space occupancy in lieu of particle density.
The dynamics of the local temperature is the only quantity we need
to model.

We now return to study the collision terms seen on
the RHS of Eq.\,(\ref{drho/dt1}). 
A related  quantity is the (L-invariant)
production  rate $A^{12\to 34}$ of particles 
per unit time and space, defined usually 
with respect to chemically equilibrated distributions: 
\begin{equation}\label{prodgen}
A^{12\to 34}\equiv\frac1{1+\delta_{1,2}} \rho_1^\infty\rho_2^\infty 
             \langle \sigma_{s} v_{12}\rangle_T^{12\to 34}  \,.
\end{equation}
The factor $1/(1+\delta_{1,2})$ is introduced to compensate 
double-counting of identical particle pairs. 
In terms of the L-invariant $A$\,,
Eq.\,(\ref{drho/dt1}) takes the form:
\begin{eqnarray}
&&\dot T\rho_{s}^\infty\left({{d\gamma_{s}}\over{dT}}+
\frac{\gamma_{s}}{T}z\frac{K_1(z)}{K_2(z)}\right)
=
\gamma_g^2(\tau)A^{gg\to s\bar s} +\nonumber\\
&&\hspace{2cm}+\gamma_{q}(\tau)\gamma_{\bar q}(\tau)A^{q\bar q\to s\bar s} 
\!-\gamma_{s}(\tau)\gamma_{\bar s}(\tau)(A^{s\bar s\to gg}
\!+A^{s\bar s\to q\bar q}).
\label{rho-gam}
\end{eqnarray}
Only weak interactions convert quark flavors, thus, 
on  hadronic time scale, we have 
$\gamma_{s,q}(\tau)=\gamma_{{\bar s},{\bar q}}(\tau)$. Moreover, 
detailed balance, arising from the time reversal symmetry of the
microscopic reactions, assures that the invariant rates
for forward/backward reactions are the same, specifically
\begin{equation}\label{det-bal}
A^{12\to 34}=A^{34\to 12},
\end{equation}
and thus:
\begin{eqnarray}
\dot T\rho_{s}^\infty\left({{d\gamma_{s}}\over{dT}}+
\frac{\gamma_{s}}{T}z\frac{K_1(z)}{K_2(z)}\right)
&=&
\gamma_g^2(\tau)A^{gg\to s\bar s}
    \left[1-\frac{\gamma_{s}^2(\tau)}{\gamma_g^2(\tau)}\right] \nonumber\\
&&
+\gamma_{q}^2(\tau)A^{q\bar q\to s\bar s}
    \left[1-\frac{\gamma_{s}^2(\tau)}{\gamma_{q}^2(\tau)}\right]\,.
\label{rho-bal}
\end{eqnarray}
When all $\gamma_i\to 1$, the Boltzmann collision term vanishes, we have
reached equilibrium.

As discussed, the gluon chemical equilibrium
is thought to be reached at high temperatures well before the strangeness
equilibrates chemically, and thus we assume this in what follows, and 
the initial conditions we will study refer to the time at which gluons
are chemically equilibrated. Setting $\lambda_g=1$ (and without a
significant further consequence for what follows, 
since gluons dominate the production rate, also $\lambda_{q}=1$)
we obtain after a straightforward manipulation the dynamical 
equation describing the evolution of the local phase space occupancy
of strangeness:
\begin{equation}\label{dgdtf}
2\tau_{s}\dot T\left({{d\gamma_{s}}\over{dT}}+
\frac{\gamma_{s}}{T}z\frac{K_1(z)}{K_2(z)}\right)
=1-\gamma_{s}^2\,.
\end{equation}
Here, we  defined the relaxation time  $\tau_{s}$  of 
chemical (strangeness) equilibration as the 
ratio of the equilibrium density that
is being approached, with the rate at which this occurs:
\begin{equation}\label{tauss}
\tau_{s}\equiv
{1\over 2}{\rho_{s}^\infty\over{
(A^{gg\to s\bar s}+A^{q\bar q\to s\bar s}+\ldots)}}\,.
\end{equation}
The factor 1/2 is introduced by convention in order for the quantity
$\tau_{s}$ to describe the exponential approach to equilibrium.

Eq.\,(\ref{dgdtf})  is our final analytical result describing 
the  evolution of phase space  occupancy.
Since one generally  expects that $\gamma_{s}\to 1$ in a monotonic 
fashion as function of time, it is important to appreciate that this 
equation allows  $\gamma_{s}>1$:
when $T$ drops below  $m_{s}$, and $1/\tau_{s}$ becomes small,
the dilution term (2nd term on LHS)
in Eq.\,(\ref{dgdtf})  dominates the evolution of $\gamma_{s}$\,.
In simple terms,  the high
abundance of strangeness produced at high temperature over-populates
the available phase space at lower temperature, when the equilibration
rate cannot keep up with the expansion cooling.
This behavior of $\gamma_{s}$  has been 
shown in \cite[Fig.\,2]{Let97} for the SPS 
conditions with fast transverse expansion. 
Since we assume that the dynamics of transverse expansion 
of QGP is similar at RHIC as at SPS, 
we will obtain a rather similar behavior for $\gamma_{s}$. 
We note that yet a faster transverse  expansion than considered here
could enhance the chemical strangeness 
anomaly.

$\tau_{s}(T)$\,, Eq.\,(\ref{tauss}),
has been evaluated using pQCD cross section and
 employing NLO (next to leading order)
running of both the strange quark mass and QCD-coupling 
constant $\alpha_{s}$ \cite{budap}. We believe that this method produces
a result for $\alpha_{s}$ that can be trusted  
down to 1\,GeV energy scale which
is here relevant. We employ results obtained with
$\alpha_{s}(M_{Z^0})=0.118$ and 
$m_{s}(\mbox{1GeV})=200$\,MeV; we have shown 
results with $m_{s}(\mbox{1GeV})=220$\,MeV earlier \cite{RL99}.
There is some systematic  uncertainty due to the appearance of the 
strange quark mass as a fixed rather than running value
in both, the chemical equilibrium 
density $\rho_{s}^\infty$ in Eq.\,(\ref{tauss}), and in the dilution term
in Eq.\,(\ref{dgdtf}). We use the value 
$m_{s}(\mbox{1\,GeV})$, with the 1\,GeV energy
scale chosen to correspond to typical interaction scale in the QGP
at temperatures under consideration.

\section{Expectations for Strange Hadron Production at RHIC}
\label{rhicsec}
We now combine our advances in theoretical models of
strangeness production and data interpretation at SPS 
energies with the objective of making reliable predictions for
the RHIC energy range \cite{RL99}. First we address the question
how much strangeness can be expected at RHIC. 
Numerical study of Eq.\,(\ref{dgdtf}) becomes possible  
as soon as we define the temporal evolution of the 
temperature for RHIC conditions.   
We expect that a global cylindrical expansion 
should describe the dynamics: aside of the longitudinal flow, we 
allow  the cylinder surface to expand given the internal thermal pressure.
SPS experience suggests that the transverse matter flow will not 
exceed the sound velocity of relativistic matter $v_\bot\simeq c/\sqrt{3}$.
We recall that for pure longitudinal expansion local entropy density scales
as $S\propto T^3\propto 1/\tau$, \cite{Bjo83}. 
It is likely that the transverse flow of matter 
 will accelerate the drop in entropy density. We thus
 consider the following  temporal evolution function 
of the temperature:
\begin{equation}\label{Toft}
T(\tau)=T_0\left[
\frac{1}{(1+\tau\ 2c/d)(1+\tau\ v_\bot/R_\bot)^2}
\right]^{1/3}\,.
\end{equation}
We take the thickness of the initial collision region at $T_0=0.5$\,GeV
to be\  $d(T_0=0.5)/2=0.75$\,fm,  and the transverse dimension in nearly 
central Au--Au  collisions to be $R_\bot=4.5$\,fm. 
The time at which thermal initial conditions are reached is
assumed to be $\tau_0=1$fm/$c$. When we vary $T_0$, 
the temperature  at which the  gluon equilibrium is reached, we also
scale the longitudinal dimension according to:
\begin{equation}\label{dofT0}
d(T_0)=(0.5\mbox{\,GeV}/T_0)^3 1.5\mbox{\,fm}\,.
\end{equation}
This assures that when comparing the different evolutions of $\gamma_{s}$ we
are looking at an initial system that has the same entropy 
content by adjusting its initial volume $V_0$. 
The reason we vary the initial temperature $T_0$
down to 300 MeV, maintaining the initial entropy content
is to understand how the assumption about the 
chemical equilibrium of gluons, reached by definition at  $T_0$,
impacts our result. In fact when considering decreasing  $T_0$  (and thus
increasing $V_0$), what we are doing is to begin the thermal production 
at a later time in the history of the collision.

The numerical integration of Eq.\,(\ref{dgdtf}) is started at 
$\tau_0$, and a range of initial temperatures $300\le T_0\le 600$, 
varying in steps of 50 MeV.   The high limit of the temperature
we  explore exceeds somewhat the `hot glue scenario' \cite{Shu92},
while the lower limit of $T_0$ corresponds to the more conservative
estimates of possible initial conditions \cite{Bjo83}.
Since the initial $p$--$p$ collisions also
produce strangeness,  we take as an estimate of initial 
abundance a common initial value
$\gamma_{s}(T_0)=0.15$. The time evolution in 
the plasma phase is followed up to
the break-up of QGP.  This condition we establish
in view of our analysis of the SPS results. We recall
that SPS-analysis showed that the system dependent 
baryon and antibaryon  $m_\bot$-slopes of particle spectra are
result of differences in collective  flow in the deconfined 
QGP source at freeze-out. There is a universality of
physical properties of hadron chemical  freeze-out between different
SPS systems, and  in our analysis a practical coincidence of 
the kinetic  freeze-out conditions with the chemical freeze-out.
 We thus expect, extrapolating the 
phase boundary curve to the small baryochemical potentials,
that the QGP break-up temperature 
$T_f^{\mbox{\scriptsize SPS}}\simeq 145\pm5$~MeV  
will see just a minor upward change 
to the value $T_f^{\mbox{\scriptsize  RHIC}}\simeq 150\pm5$~MeV.

\begin{figure}[t]
\vspace*{1.4cm}
\centerline{\hspace*{-0.3cm}\psfig{width=7.4cm,figure=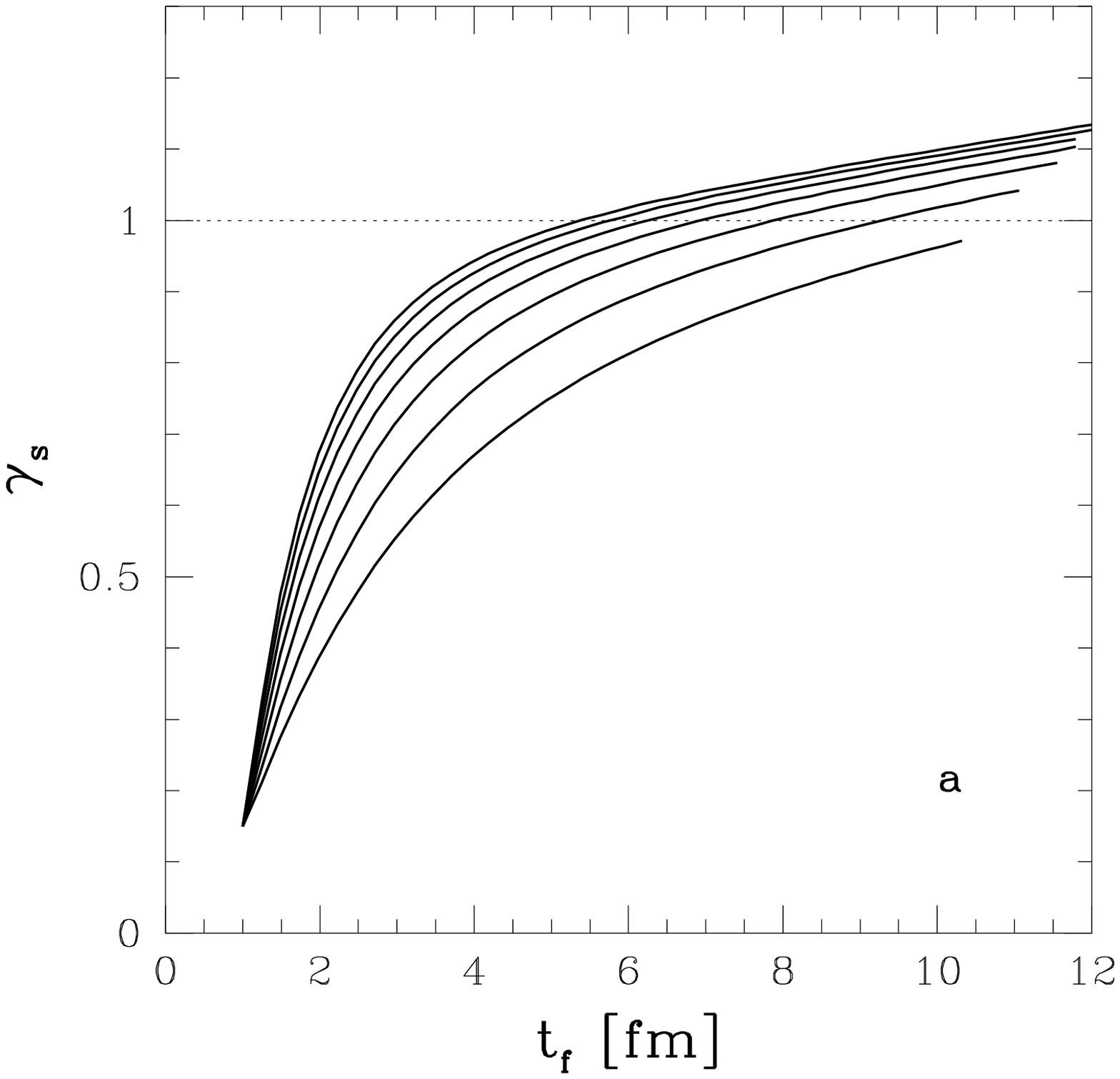}
\hspace*{-1.2cm}\psfig{width=7.4cm,figure=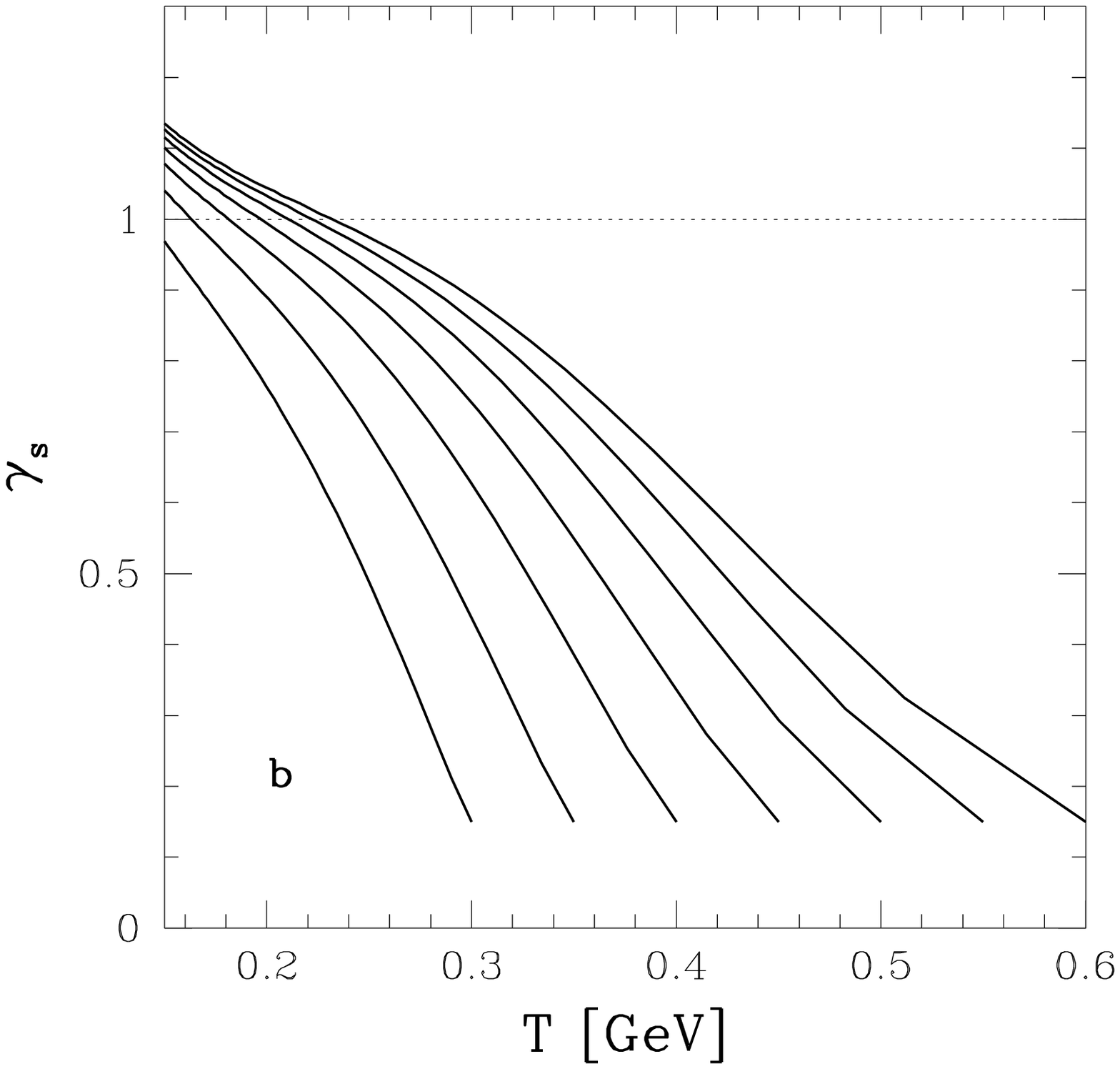}}
\vspace*{-0cm}
\caption{ 
Evolution of QGP-phase strangeness phase space occupancy $\gamma_{s}$. 
{\bf a}) as function of  time and, {\bf b}) as function
of temperature for $m_s(1\,\mbox{GeV})=200$\,MeV, see text for details.
\label{figvarsgam}} 
\end{figure}

With the freeze-out condition fixed, one would think that the major uncertainty 
in our approach comes from the initial gluon equilibration 
temperature $T_0$, and we now study how different values of $T_0$ influence
the final state phase space occupancy. We integrate numerically Eq.\,(\ref{dgdtf}) 
and present $\gamma_{s}$ as function of both time $t$ in
Fig.\,\ref{figvarsgam}a, and temperature $T$ in Fig.\,\ref{figvarsgam}b, 
up to the expected QGP breakup at 
$T_f^{\mbox{\scriptsize RHIC}}\simeq 150\pm5$ MeV. 
 We see that: \\
$\bullet$ widely different initial
conditions (with similar initial entropy content) 
lead to rather similar chemical conditions at chemical freeze-out 
of strangeness, \\
$\bullet$  despite
a series of conservative assumptions, we find, not only, that strange\-ness
equilibrates, but indeed that the dilution effect allows an overpopulation of 
the strange quark phase space.
 For a wide range of initial conditions, we obtain a 
narrow band $1.15>\gamma_{s}(T_f)>1$\,.  
We will in the following, taking into account some contribution from hadronization
of gluons in strange/antistrange quarks,  adopt what the value $\gamma_{s}(T_f)=1.25$. 

We now consider how this relatively large value of $\gamma_{s}$,
characteristic for the underlying QGP formation and evolution 
of strangeness, impacts the strange baryon and anti-baryon observable
emerging in hadronization.  Remembering that major changes compared to 
SPS should occur in rapidity 
spectra of mesons, baryons and antibaryons, 
we will apply the same hadronization model that 
worked in the analysis of the SPS data.
This hypothesis can be falsified 
easily, since we expect, based and  compared to the 
Pb--Pb 158$A$ GeV results:\\
\indent a) shape identity  
of all RHIC $m_\bot$ and $y$ spectra of antibaryons 
 $\bar p\,,\ \overline\Lambda\,,\ \overline\Xi\,,$ since 
in our approach there is no difference in 
their production mechanism, and the form of the spectra is determined
in a similar way by the local temperature and flow velocity vector; \\
\indent b) the $m_\bot$-slopes of these antibaryons
 should be very similar to the result we
have from  Pb--Pb 158$A$ GeV
since only a slight increase in the freeze-out temperature 
occurs, and no increase in collective transverse flow is expected.

The abundances of particles produced from QGP  
within the sudden  freeze-out model are controlled by several 
parameters we addressed earlier: the light quark fugacity 
$1<\lambda_{q} <1.1$\,,  value is limited by the expected small
ratio between baryons and mesons (baryon-poor plasma) when the 
energy per baryon is above 100\,GeV, strangeness fugacity 
$ \lambda_{s}\simeq 1$ 
which value for locally neutral plasma assures that 
$\langle s-\bar s\rangle =0$; 
the light quark phase space occupancy 
$\gamma_{q}\simeq 1.5$, overabundance value due to  gluon fragmentation. 
Given these narrow ranges of chemical parameters and 
the freeze-out temperature $T_f=150$ MeV, 
 we compute the expected particle production at break-up. 
In general we cannot expect that the absolute numbers of particles we
find are correct, as we have not modeled the important effect of flow in
the laboratory frame of reference. However, ratios of
hadrons subject to similar flow effects (compatible hadrons)
can be independent of  the detailed final state dynamics, as 
the results seen at SPS suggest, and
we will look at such ratios more closely.

Taking $\gamma_{q}=1.5{\tiny\begin{array}{c}+0.10\\-0.25 \end{array}}$ 
we choose the value of 
$\lambda_{q}$, see the header  of table \ref{table1}, 
for which the  energy  per baryon ($E/B$)
is similar to the collision condition 
(100\,GeV),  which leads to the 
range $\lambda_{q}=1.03\pm0.005$. We  evaluate for 
these  examples aside of $E/B$, the strangeness per baryon 
$s/B$ and entropy per baryon $S/B$ as shown in the top 
section of the table \ref{table1}. We do
not enforce  $\langle s-\bar s\rangle=0$ exactly, but 
since baryon asymmetry is
small, strangeness is  balanced to better than 2\%\, in
the parameter range considered.
In the bottom portion of  table \ref{table1},  we present
the compatible particle abundance ratios,
computed according to the procedure developed 
in section \ref{fermisec}.  We have given,
aside of the baryon and antibaryon relative yields, also the relative 
kaon yield, which is also well determined within our approach. 
\begin{table}[t]
\caption{\label{table1} 
For $\gamma_{s}=1.25,\,\lambda_{s}=1$ and $\gamma_{q}$, $\lambda_{q}$ as shown:
Top portion: strangeness per baryon $s/B$, 
energy per baryon $E/B$[GeV]  and  entropy per baryon $S/B$. Bottom portion:
sample of hadron ratios expected at RHIC.}
\small
\vspace*{-0.2cm}
\begin{center}
\begin{tabular}{l|lllll}
\hline\hline
 $\gamma_{q}$                               & 1.25 & 1.5  &  1.5  &  1.5  & 1.60 \\
$\lambda_{q}$                               & 1.03 & 1.025&  1.03 & 1.035 & 1.03 \\
\hline\hline
$E/B$[{\small GeV}]                       & 117  & 133  &  111  &  95   & 110 \\
$s/B$                                     & 18   & 16   &  13   &  12   & 12 \\
$S/B$                                     & 630  & 698  &  583  & 501   & 571 \\
\hline
$p/{\bar p}$                              & 1.19 & 1.15 & 1.19  &  1.22 & 1.19 \\
$\Lambda/p$                               & 1.74 & 1.47 & 1.47  &  1.45 & 1.35 \\
${\bar\Lambda}/{\bar p}$                  & 1.85 & 1.54 & 1.55  &  1.55 &1.44 \\
${\bar\Lambda}/{\Lambda}$                 & 0.89 & 0.91 &  0.89 &  0.87 & 0.89 \\
${\Xi^-}/{\Lambda}$                         & 0.19 & 0.16&  0.16  &  0.16 & 0.15 \\
${\overline{\Xi^-}}/{\bar\Lambda}$          & 0.20 & 0.17 &  0.17 &  0.17 & 0.16 \\
${\overline{\Xi}}/{\Xi}$                  & 0.94 & 0.95 &  0.94 &  0.93 & 0.94 \\
${\Omega}/{\Xi^-}$                                    & 0.147&0.123 &  0.122&  0.122& 0.115 \\
${\overline{\Omega}}/{\overline{\Xi^-}}$              & 0.156& 0.130&  0.130&  0.131& 0.122 \\
${\overline{\Omega}}/{\Omega}$                      &  1   & 1.   &  1.   &  1.   & 1.   \\
$\Omega+\overline{\Omega}\over\Xi^-+\overline{\Xi^-}$    & 0.15 & 0.13 &  0.13 &  0.13 & 0.12 \\
$\Xi^-+\overline{\Xi^-}\over\Lambda+\bar{\Lambda}$       & 0.19 & 0.16 &  0.16 &  0.16 & 0.15 \\
${K^+}/{K^-}$                                       & 1.05 & 1.04 &  1.05 &  1.06 & 1.05 \\
\end{tabular}
\end{center}
\vspace*{-.6cm} 
\end{table}
\begin{table}[htb]
\caption{\label{table2} $dN/dy|_{\mbox{\scriptsize cent.}}$ 
assuming in this example $dp/dy|_{\mbox{\scriptsize cent.}}=25$ .}
\vspace*{-.2cm} 
\begin{center}
\begin{tabular}{ll|cccccccccc}
\hline\hline
$\gamma_{q}$& $\lambda_{q}$  & $b$ &  $p$ & $\bar p$ & $\!\!\Lambda\!\!+\!\!\Sigma^0\!\!$ & $\!\!\overline{\Lambda}\!\!+\!\!\overline{\Sigma}^0\!\!$&$\Sigma^{\pm}$&$\overline{\Sigma}^{\mp}$ & 
$\Xi^{^{\underline{0}}}$ &$\overline{\Xi}^{^{\underline{0}}}$& $\Omega\!=\!\overline\Omega$  \\
\hline
1.25& 1.03 & 17 & 25$^*$& 21 & 44 & 39 & 31 & 27 & 17 & 16 & 1.2  \\ 
1.5 & 1.025 & 13 & 25$^*$& 22 & 36 & 33 & 26 & 23 & 13 & 11 & 0.7 \\ 
1.5& 1.03 & 16 & 25$^*$& 21 & 37 & 33 & 26 & 23 & 12 & 11 & 0.7 \\ 
1.5 & 1.035 & 18 & 25$^*$& 21 & 36 & 32 & 26 & 22 & 11 & 10 & 0.7 \\ 
1.60& 1.03  & 15 & 25$^*$& 21 & 34 & 30 &24 &21 & 10 & 9.6 & 0.6 \\ 
\end{tabular}
\end{center}
\vspace*{-0.6cm} 
\end{table}

The meaning of these results can be better appreciated when
we assume in an example the central  rapidity density
of direct protons is  $dp/dy|_{\mbox{\scriptsize cent.}}=25$. 
In table  \ref{table2},  we present the 
resulting (anti)baryon abundances. 
We see that the  net baryon density $db/dy\simeq 16\pm3$,  there is 
baryon number transparency. We see that (anti)hyperons are 
indeed more abundant than non-strange  (anti)baryons.
Taking into account the disintegration of strange baryons,
we are finding a much greater  number of observed protons 
$dp/dy|_{\mbox{\scriptsize cent.}}^{\mbox{\scriptsize obs.}}
\simeq 65\pm5$ in the central rapidity region. It is important
when quoting results from table  \ref{table2} to recall that:\\
\indent 1) we have chosen arbitrarily the overall
normalization in table  \ref{table2}\,, only particle ratios 
were computed,  and\\ 
\indent 2) the rapidity baryon density
relation to rapidity proton density is a consequence of the assumed 
value of $\lambda_{q}$, which we chose to get 
$E/B\simeq 100$\,GeV per  participant. 

The most interesting result seen in table  \ref{table2}\,, 
the hyperon-dominance of the baryon yields at RHIC, does not depend on detailed 
model hypothesis. We have  explored another set of parameters in
our first and preliminary report on this matter \cite{prelim}, finding this
result. Another interesting property of the hadronizing hot RHIC
matter as seen in  table \ref{table1}, is that strangeness yield
per participant is expected to be 13--23 times greater than seen at present at 
SPS energies, where we have 0.75 strange quark pairs per baryon.
As seen in  table  \ref{table2}, the baryon rapidity density is in our examples
similar to the proton rapidity density.

\section{Conclusions}
We believe that the Fermi model interpretation of SPS 
strangeness results decisively shows some interesting new 
physics. We see considerable convergence of the results 
around properties of suddenly hadronizing QGP. 
The key results we obtained  are: \\
\indent 1) same hadronization temperature $T$=142--144\,MeV 
for very different collision systems with different hadron spectra;\\
\indent 2) QGP expected $\tilde \lambda_{s}=1$  for
S and Pb collisions, and  $\lambda_{s}^{\rm Pb}\simeq 1.1$\,;\\ 
\indent 3) $\gamma_{s}^{\rm Pb}>1$, indicating that high strangeness yield was reached before
freeze-out;\\
\indent 4) $\gamma_{q}>1$\, as would be expected from high entropy phase and the associated
value  $S/B\simeq 40$\,;\\
\indent 5) yield of strangeness per baryon $\bar s/B\simeq 0.75$ just as predicted by gluon  
fusion in thermal QGP;\\
\indent 6) transverse expansion velocity $v_c^{\rm Pb}=1/\sqrt{3}$, the sound
velocity of quark matter for Pb--Pb. 

Among other interesting results which also verify the consistency
of our approach,  we recall:\\ 
\noindent$\bullet$ the  exact balancing of
strangeness $\langle \bar s-s\rangle=0$ in the symmetric Pb--Pb case;\\ 
\noindent$\bullet$ 
increase of the baryochemical potential 
$\mu_B^{\rm Pb}={203\pm5} >\mu_B^{\rm S}=178\pm5$\,MeV
as the collision system grows;\\ 
\noindent$\bullet$ 
energy per baryon near to the value expected if energy and baryon number
deposition in the fireball are similar.\\
The universality of the physical properties
at chemical freeze-out for S- and Pb-induced
reactions points to a common nature of the
primordial source  of hadronic particles. The difference
in spectra between the two systems arises in our analysis due to the difference
in the collective surface explosion velocity, 
$v_c^{\rm S}=0.5\,<\,v_c^{\rm Pb}=1/\sqrt{3}$\,,
which for larger system  is higher, having  more time to develop.

In our opinion,  these results show 
that hadronic particles seen  at CERN-SPS
are  emerging from a deconfined QGP phase of hadronic matter
and do not undergo a  re-equilibration after they have been
produced.  This has encouraged us to consider within the same
computational scheme the production of strange hadrons at RHIC 
conditions, and  we have shown that one can expect strangeness chemical 
equilibration in nuclear collisions at RHIC if
the deconfined  QGP is formed, with a probable overpopulation 
effect associated with the early strangeness abundance freeze-out
before hadronization. We have shown also that  (anti)hyperons dominate 
(anti)baryon abundance, and that rapidity distributions of
(anti)protons are primarily deriving from decays of (anti)hyperons.





\end{document}